\documentclass[journal]{IEEEtran}
\usepackage[a4paper, margin=2.7cm]{geometry}

\usepackage[pdftex]{graphicx}
\usepackage{epstopdf}

\usepackage{amsmath, amsthm, amsfonts, amssymb}
\usepackage{mathrsfs}

\usepackage{etoolbox,lineno}
\usepackage{authblk}

\usepackage{xcolor}
\usepackage{hyperref}
\usepackage{subcaption}
\usepackage{graphicx}
\usepackage{tikz}
\usepackage{calc}

\title{Adaptive Optimal Control for Avatar-Guided Motor Rehabilitation in Virtual Reality}

\begin{document}

\author[1,2]{}
\author[1,2]{}

\author[1,2,*]{Francesco De Lellis\textsuperscript{1,$\dagger$}, Maria Lombardi\textsuperscript{2,$\dagger$}, Egidio De Benedetto\textsuperscript{1,3}, Pasquale Arpaia\textsuperscript{1,3}, Mario di Bernardo\textsuperscript{1,3,4,*}
\thanks{\textsuperscript{1}Department of Electrical Engineering and
Information Technology, University of Naples Federico II, Naples, Italy.}%
\thanks{\textsuperscript{2}Humanoid Sensing and Perception, Istituto Italiano di Tecnologia, Genoa, Italy.}%
\thanks{\textsuperscript{3}Interdepartmental Research Center on Management and Innovation in Healthcare, University of Naples Federico II, Naples, Italy.}%
\thanks{\textsuperscript{4}Scuola Superiore Meridionale, Naples, Italy.}%
\thanks{This work was financially supported by the Italian Ministry of Health, through the project HubLife Science-Digital Health (LSH-DH) PNC-E3-2022-23683267- DHEAL COM – CUPE63C22003790001, within the 'National Plan for Complementary Investments—Innovative Health Ecosystem' — Unique Investment Code: PNC-E.3. This publication reflects only the authors’ view and the Italian Ministry of Health is not responsible for any use that may be made of the information it contains.
}%
\thanks{\textsuperscript{$\dagger$}These authors contribuited equally.}%
\thanks{\textsuperscript{*}Corresponding author: mario.dibernardo@unina.it}%
}

\maketitle


\begin{abstract}
\noindent A control-theoretic framework for autonomous avatar-guided rehabilitation in virtual reality, based on interpretable, adaptive motor guidance through optimal control, is presented. 
The framework faces critical challenges in motor rehabilitation due to accessibility, cost, and continuity of care, with over 50\% of patients inability to attend regular clinic sessions. 
The system enables post-stroke patients to undergo personalized therapy in immersive virtual reality at home, while being monitored by clinicians. 
The core is a nonlinear, human-in-the-loop control strategy, where the avatar adapts in real time to the patient's performance. 
Balance between following the patient’s movements and guiding them to ideal kinematic profiles based on the Hogan minimum-jerk model is achieved through multi-objective optimal control.
A data-driven "ability index" uses smoothness metrics to dynamically adjust control gains according to the patient's progress. 
The system was validated through simulations and preliminary trials, and shows potential for delivering adaptive, engaging and scalable remote physiotherapy guided by interpretable control-theoretic principles.
\end{abstract}

\section{Introduction}
\label{sec:intro}

Home-based physiotherapy supported by virtual reality (VR) technologies is emerging as a promising alternative to traditional rehabilitation models, particularly in post-injury and post-stroke care. By combining immersive environments with motion tracking and feedback systems, VR-based approaches improve accessibility for patients with mobility limitations or living in remote areas, while reducing the need for frequent in-person visits. Recent studies estimate that over half of individuals requiring physiotherapy cannot attend in-clinic sessions regularly, highlighting the need for scalable and remote solutions \cite{reilly2021virtual}. Structured VR platforms have the potential to deliver consistent therapeutic experiences and enable real-time progress monitoring by clinicians \cite{corbetta2015rehabilitation, brepohl2023virtual}.
Modern VR systems often incorporate advanced motion-tracking technologies to monitor kinematic parameters and provide real-time audiovisual feedback. These tools can be further enhanced by autonomous virtual agents (avatars) that interact with users to simulate therapeutic guidance. The immersive nature of VR has also been shown to reduce pain perception during physically demanding exercises through sensory distraction \cite{morris2009effectiveness}. Within this context, cognitive architectures play a central role in generating intelligent avatar behavior by coordinating complex information flows from motion data. Prior approaches based on black-box machine learning have demonstrated avatar-human synchronization capabilities \cite{lombardi2021using, GROTTA202437}, but often lack interpretability and require extensive training data, limiting their clinical deployability \cite{delellis_datadriven}.

In this study, we focus on stroke rehabilitation as an illustrative scenario and on the design of interpretable avatar behaviors powered by adaptive optimal control strategies to interact with human patients and extract key principles of physiotherapy processes. 
We propose a novel VR-based rehabilitation platform featuring an autonomous virtual avatar that supports the patient throughout the therapy process by adapting its behavior to construct personalized therapy sessions. 
Powered by our control solutions, the artificial avatar provides adaptive visual feedback in real-time through co-execution of motor tasks, facilitating engagement and correction, and providing a crucial element to achieve positive results in rehabilitation programs \cite{hamzeheinejad2021impact}.
Furthermore, our proposed architecture integrates analytical tools based to process the continuous stream of motion data captured by the sensor system for three purposes: i) assessing the patient's level of impairment, ii) monitoring their progression, and iii) informing therapy plan personalization based on quantitative metrics.

\begin{figure*}[t]
    \centering

    \hspace{1em}
    \begin{subfigure}[b]{0.55\textwidth}
        \centering
        \includegraphics[width=.9\textwidth]{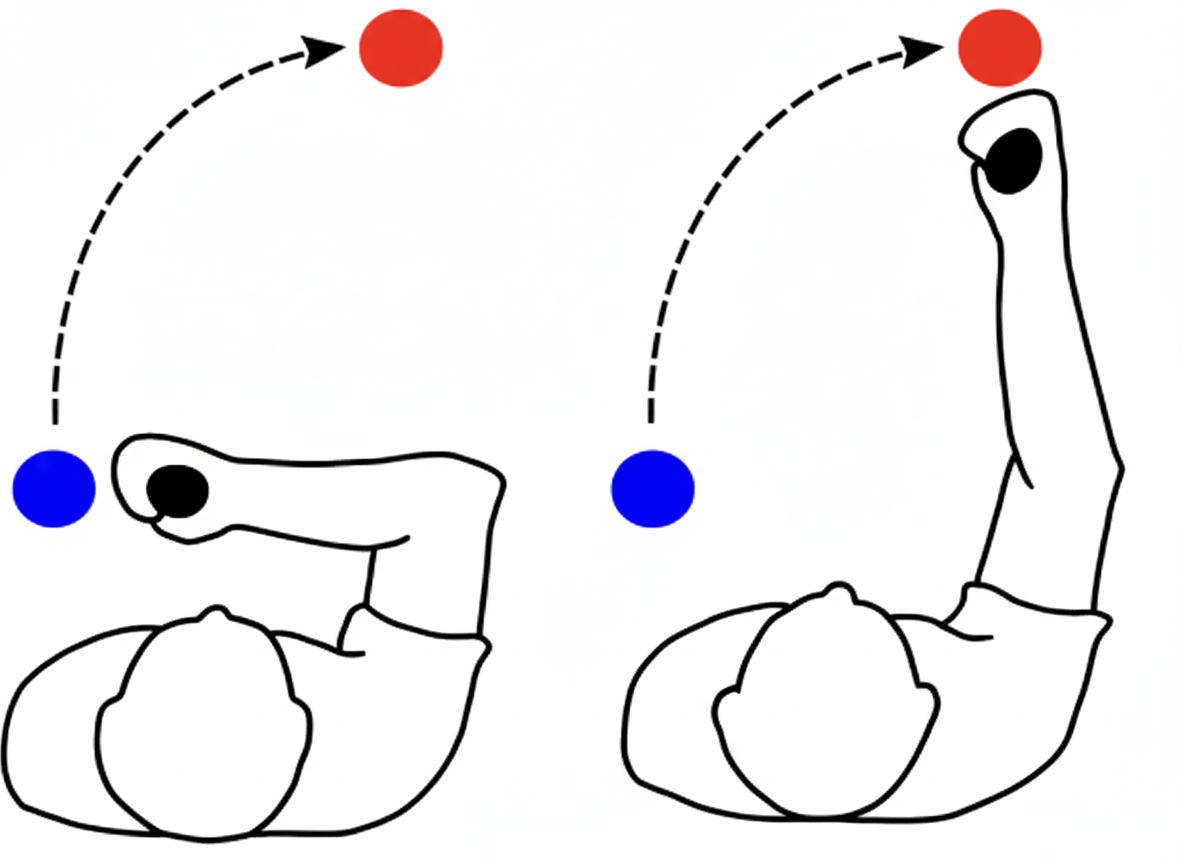}
        \caption{}
        \label{fig:illustration_a}
    \end{subfigure}
    \hfill
    \begin{subfigure}[b]{0.35\textwidth}
        \centering
        \includegraphics[width=.9\textwidth]{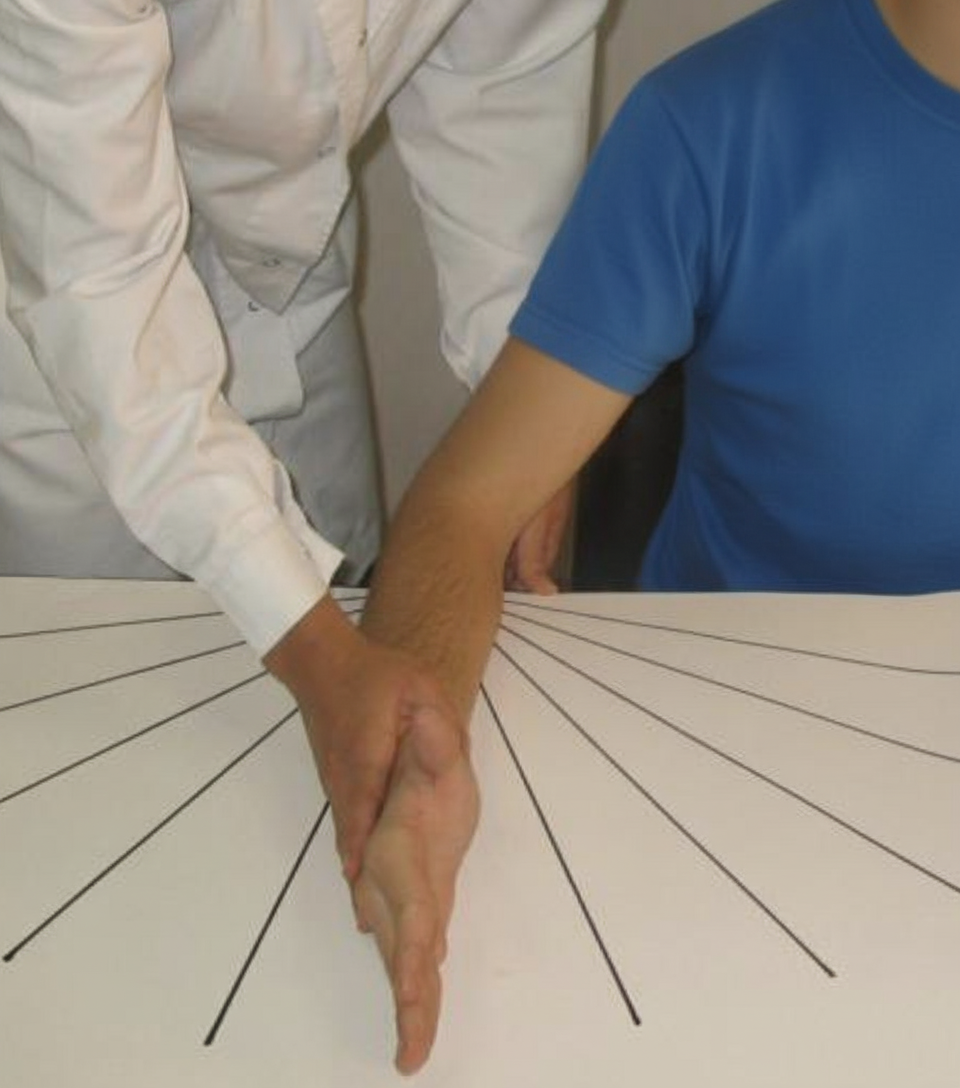}
        \caption{}
        \label{fig:illustration_b}
    \end{subfigure}
    \hspace{1em}

	\caption{\textbf{Motor task used in the design of the rehabilitation platform.} \textbf{(a)} Illustration representing a human performing the reaching task at the beginning and at the end of the task. Blue circle represents the start angle position, whereas the red circle represents the final angle position. A dashed arrow indicates the trajectory that the human has to follow. \textbf{(b)} Real scenario in which human is asked to move the forearm in an angular motion assisted by the clinician.}
	\label{fig:motor_task}
\end{figure*}

\section{Related Works}
Stroke remains a leading cause of long-term motor disability worldwide, with hemiparesis of the contralateral upper limb being the most common impairment \cite{Feigin2009, Thrift2014}. These motor deficits hinder daily activities such as reaching, grasping, and holding, significantly reducing quality of life \cite{Cramer1997, Hatem2016}.

In the standard healthcare model, post-stroke rehabilitation involves daily, systematic exercises performed by the patient under the direct supervision of clinicians. This process typically includes two phases: i) an initial \emph{acute phase}, where initial rehabilitation occurs during hospitalization, followed by ii) a \emph{long-term outpatient therapy} aimed at restoring motor functions through repeated supervised exercises. However, due to increasing healthcare costs and resource limitations, access to continuous outpatient support is becoming more difficult to sustain \cite{WHO_ScaleUpRehab_2018}. These limitations motivate the development of rehabilitation platforms capable of delivering high-quality therapy outside clinical environments.

VR-based rehabilitation platforms have been proposed as viable tools to extend care beyond hospital settings \cite{leite2025use, Laver11}, moreover, several clinical studies reported positive effects of specific VR physiotherapy applications \cite{Fu2015, de2016effect}. Numerous VR-based rehabilitation systems have been developed to support post-stroke motor recovery through gamified exercises targeting various motor skills. For instance, games such as ``Ball in Hole'', ``Cloud, Glasses'', and ``Rolling Pin'' require users to execute precise hand movements by manipulating virtual objects via gesture control \cite{sramka2020,fregna2022}. Similarly, a ``Tetris''-inspired game developed by \cite{ferreira2020} promotes hand coordination as patients control falling shapes.
For more complex activities, games like ``Balloon Blast'' \cite{bailey2022} engage users in shoulder mobilization through swiping gestures, and ``Rolling Pin'' \cite{fregna2022} promotes bilateral gross motor skills through coordinated two-handed movements.
Moreover, access to modern day tracking systems for VR could also allow intense AI-driven processing of motions for stroke patient to inform on current physiotherapy progress and future planning \cite{trabassi2025}. 

While these systems include some mechanisms for task adaptation based on user performance, they generally lack an interactive partner. 
Moreover, the training parameters are not automatically adjusted in real time based on evolving individual subject  performance. For example, in \cite{sramka2020,bailey2022} adaptation was manually implemented by the clinician through direct supervision and feedback. In unsupervised scenarios \cite{ferreira2020}, adaptation relied on post-session analysis of performance metrics, which required clinicians to later update training content manually. This limits the systems' capacity for autonomous and responsive personalization during the rehabilitation process.

Finally, recent literature findings strongly emphasize the need for modern systems that can dynamically adjust to patient performance while providing interpretable metrics for clinicians \cite{kuhne2024virtual, hao2023effects}.

\section{Design of the rehabilitation platform}
\label{sec:rehabilitation_platform}
To allow effective real-time human interaction during motor tasks, we present the fundamental principles behind our rehabilitation platform. 

\subsection{Reaching motor task}
\label{sec:reaching_task}
Reaching tasks are representative motor exercises frequently employed in clinical stroke rehabilitation \cite{Collins2018}. The specific task studied here involves a seated patient, with the elbow fixed as a pivot, moving their forearm along a circular arc between target points as schematically depicted in Fig. (\ref{fig:motor_task}). 

Reaching tasks are also valuable in research to understand the kinematic and dynamic aspects of voluntary arm movements. In fact, voluntary point-to-point arm movements from healthy subjects tend to follow smooth trajectories with a single-peaked, bell-shaped velocity profile \cite{Soechting1981, Georgopoulos1981}.

In his seminal work, Neville Hogan proposed a mathematical model (the \emph{Hogan model}) that captures the principles underlying voluntary arm movements \cite{Hogan1984}. The model posits that the smoothness observed in such movements can be explained by assuming the motor system aims to minimize the mean squared jerk:
\begin{equation}
    C = \dfrac{1}{2} \int_{0}^{T} \left( \dddot{\theta}(t) \right)^2 dt, 
\label{eq:cost_function_hogan}
\end{equation}
where $\dddot{\theta}(t)$ represents the angular jerk profile over the movement duration $T$. Minimizing this jerk is typically subject to boundary conditions specifying the start ($\theta_0$) and final ($\theta_f$) angular positions, assuming zero initial and final angular velocities:
\begin{equation}
    \begin{cases}
        \theta(0) = \theta_0,\\
        \theta(T) = \theta_f,\\
        \dot{\theta}(0) = 0,\\
        \dot{\theta}(T) = 0,
    \end{cases}
\label{eq:boundary_conditions_hogan}
\end{equation}

Hogan demonstrated that the unique trajectory $\theta(t)$ minimizing the cost function (\ref{eq:cost_function_hogan}) subject to the boundary conditions (\ref{eq:boundary_conditions_hogan}) is a fifth-order polynomial in time:
\begin{equation}
\label{eq:optimal_trajectory_hogan}
\begin{split}
    \theta(t) = \theta_0 + (\theta_f - \theta_0) \Big[ & 10 \left(\frac{t}{T}\right)^3 - 15 \left(\frac{t}{T}\right)^4 \\
    & + 6 \left(\frac{t}{T}\right)^5 \Big] \qquad 0 \leq t \leq T.
\end{split}
\end{equation}

Notably, this trajectory's form is independent of the physical system (e.g., limb inertia). Since it effectively captures the smoothness characteristic of human movement, the \emph{Hogan model} is widely used in rehabilitation medicine as a benchmark for healthy arm trajectories. Consistent with this practice, we employ Eq. (\ref{eq:optimal_trajectory_hogan}) as the optimal reference trajectory during therapy sessions.

\subsection{Avatar-assisted therapeutic process}
The therapeutic process, facilitated by our artificial avatar, begins with a \emph{clinician-supervised assessment phase}. Its objective is to determine the patient's \emph{Range of Motion} (ROM) and gather \emph{Anthropometric Data}. To achieve this, the patient performs a reaching task multiple times without assistance, aiming for their maximum comfortable elbow extension and flexion. This data defines the patient's \emph{active ROM} as the maximum angular range (in degrees) through which the elbow joint can move. Subsequently, physical parameters of the patient's forearm (e.g., inertia, friction, and stiffness) are estimated using the recorded anthropometric data to tune avatar's internal dynamic model and enable a personalized physiotherapy session.

Following this preliminary phase, the core therapy consists of repeated sessions, each comprising two main stages: \textbf{Avatar Following (Awing)} and \textbf{Patient Following (Pawing)}.

During the \textbf{Awing phase}, the patient performs reaching tasks within their assessed ROM. Each task is repeated several times (e.g., three times by default) without direct avatar assistance. The avatar records the patient's actual movement kinematics and uses this data to update its internal mathematical model, representing the patient's current movement characteristics and calculating the corresponding optimal trajectory. The patient's task execution is then compared with the \emph{Hogan model} (\ref{eq:optimal_trajectory_hogan}) using specific performance metrics (e.g., comparing velocity profiles). This step is crucial for quantifying the patient's current performance level and identifying areas for improvement.

After collecting performance metrics, the therapy transitions into the \textbf{Pawing phase}. Here, the avatar actively demonstrates the desired smooth movement trajectory, derived from the \emph{Hogan model} or adapted based on the patient's ability. This virtual trainer provides visual, adaptive guidance to lead the patient through the exercise. This behavior encourages the patient to match the avatar's movement as closely as possible during repetitions of the same exercise (e.g., three times by default), as determined by the clinician.

These \emph{two-stage therapeutic sessions} are repeated iteratively throughout the rehabilitation program. The overarching goal is to progressively reduce the discrepancy between the patient's actual movement kinematics (particularly the velocity profile) and the smooth, single-peaked velocity profile characteristic of the \emph{Hogan model} reference trajectory for the same task. As therapy progresses, it is expected that the patient's velocity profile, which may initially exhibit multiple peaks or irregularities, will converge towards the smoother, single-peaked reference, indicating improved motor control and coordination.

The application also evaluates performance metrics in order to assess the current degree of the patient's disabilities, compares the patient's motion with the ``healthy'' trajectory given by the \emph{Hogan model}.

\section{Control Architecture for Avatar-assisted Physiotherapy}
\label{sec:architecture_avatar}

\begin{figure*}[t!]
	\centering
	\includegraphics[width=0.9\textwidth]{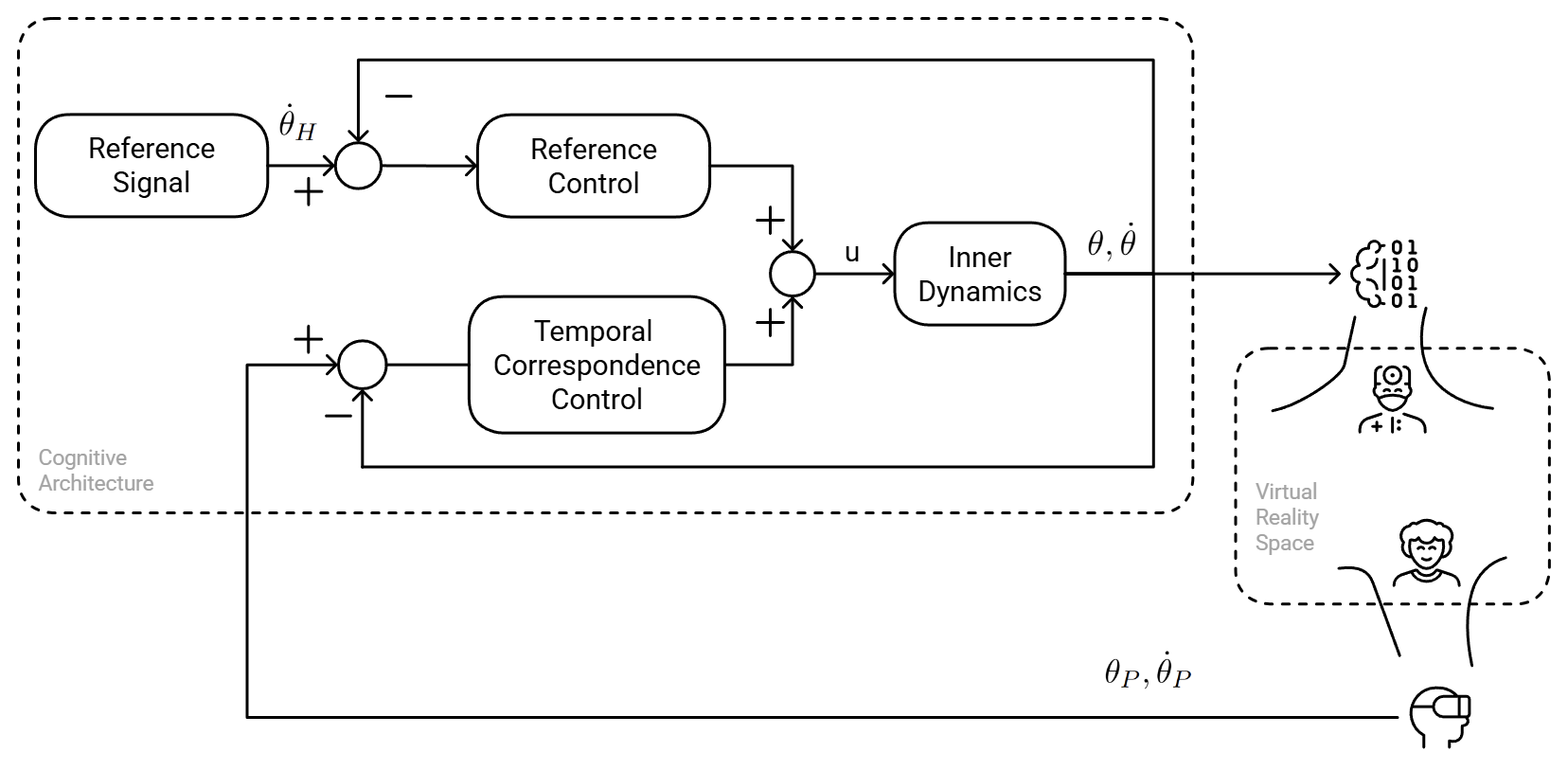}
	\caption{\textbf{Control architecture designed for the assisting avatar.} The Inner Dynamics block constraints the behavior of avatar, while the the Temporal Correspondence Control block minimizes the position error between the avatar and the human, The Reference Signal and the Reference Control blocks confer a desired kinematic features to the virtual agent. $\theta,\dot{\theta}$ are angular position and angular velocity of the avatar, whereas $\theta_P, \dot{\theta}_P$ are measured angular position and estimated angular velocity of the human partner; $\dot{\theta}_H$ is the angular velocity signal reference generated by the \emph{Hogan model}; $u$ is the control input.}
	\label{fig:cognitive_architecture_avatar}
\end{figure*}

In this section, we outline the control architecture driving the cognitive process of the avatar, designed to implement the rehabilitation strategy described earlier and support the patient during their recovery. Specifically, the avatar needs to dynamically adjust its motor behavior to achieve two primary objectives: i) guide the patient's movements towards a smooth, healthy movement pattern, and ii) adapt its own motion (for example, by slowing down or stopping or speeding up) in order to synchronize with the patient's motion to maintain engagement and achieve task completion. To meet these objectives, the control problem is formulated as a finite-horizon optimal control task embedded in a feedback loop enhanced by performance-based adaptation of the cost function.

The control architecture, illustrated in Fig. (\ref{fig:cognitive_architecture_avatar}), comprises four interconnected components that collectively govern the avatar’s behavior during rehabilitation. At the core lies the \textit{Inner Dynamics Block}, which models the avatar’s movement when not influenced by patient interaction. 
Complementing this, the \textit{Reference Signal Generator} produces a target velocity trajectory based on the minimum-jerk model proposed by Hogan~\cite{Hogan1984}. This reference serves as a normative benchmark for healthy movement and provides a goal toward which the avatar can guide the patient.
The \textit{Control Strategy} integrates two submodules: one responsible for ensuring temporal and spatial alignment with the patient’s current motion, and another that enforces convergence toward the reference trajectory. These submodules work in tandem to generate control inputs that balance synchronization with guidance, enabling a continuum of behavior from follower to leader depending on rehabilitation goals.

Finally, the architecture incorporates an \textit{ability index} calculator that quantifies the patient’s performance by comparing the executed movement with the theoretical reference. This index informs session-to-session adjustments in the control strategy, allowing the system to adapt the level of assistance to the patient’s evolving motor capabilities.
Moreover, by monitoring this index, a clinician can leverage the interpretability of our proposed solution and inform and modify the therapy plan. For example, a consistently low \textit{ability index}, despite the avatar's guidance, can be interpreted as a sign that the patient is struggling with the exercise. This could lead the clinician to ease the task, i.e., by reducing the assigned ROM to avoid a performance ceiling.

\subsection{Avatar dynamics model}
The avatar's motion is modeled using a linear harmonic oscillator that describes the angular position of its end-effector. The model is represented by the following equation:
\begin{equation}
    I\ddot{\theta} + B\dot{\theta} + K\theta = u \qquad 0 \leq \theta \leq \theta_{ROM},
\label{eq:inner_dynamics}
\end{equation}
where $\theta$ is the forearm angular position, $\dot{\theta}$ is the forearm angular velocity, $\ddot{\theta}$ is the forearm angular acceleration, $\theta_{ROM}$ is the maximum range of motion allowed, $I$ is the forearm inertia, while $B$ and $K$ represent friction and stiffness, respectively (these are assessed specifically for each patient by clinicians during the initial phase of rehabilitation). Lastly, $u$ is the time-varying control input to the system, modeling the physical inputs from the alpha neurons responsible for muscle contraction. Without loss of generality, we proceed with the control design using a damped harmonic oscillator inner dynamics due to its simplicity and effectiveness \cite{Hogan1984}. 

\subsection{Reference trajectory generator}
In addition to the measured position and velocity of the patient, the avatar receives a reference signal representing the optimal velocity profile to guide the patient's movement. This signal, derived from the \emph{Hogan model} described in Section \ref{sec:reaching_task}, represents the theoretical velocity profile of a healthy human performing a reaching movement between two points.

By differentiating equation (\ref{eq:optimal_trajectory_hogan}), we obtain the velocity reference signal:
\begin{equation}
    \dot{\theta}_H\left(t\right) = \dfrac{1}{T^5} \left[30 \, t^2 \left(\theta_f - \theta_0 \right) \left(T-t \right)^2 \right] \quad 0 \leq t \leq T,
\end{equation}
where $\theta_0$ and $\theta_f$ are the initial and final angular positions and $T$ is the total movement duration.

\subsection{Control strategy}
The time-varying input $u$ fed to the dynamics described in (\ref{eq:inner_dynamics}) is determined by a multi-objective optimal control law evaluated over a sequence of finite time intervals $[t_k, t_{k+1}]$. The cost function considers both the positional difference between the avatar and the patient, and the deviation between the avatar's velocity profile and the desired reference velocity profile. Formally, the control of the avatar, subject to the dynamics in (\ref{eq:inner_dynamics}), is formulated as the following optimization problem:
\begin{subequations}
\begin{align}
    \min_{u \in \mathbb{R}} \ & \frac{1}{2} \alpha_p \left( \theta(t_{k+1}) - \theta_P(t_{k+1}) \right)^2 \nonumber \\
    & + \frac{1}{2} \int_{t_k}^{t_{k+1}} \alpha_s \left(\dot{\theta}(\tau) - \dot{\theta}_H(\tau) \right)^2 + \eta u(\tau)^2 \; d\tau, \label{eq:prob_cost} \\
    \text{s.t.}
    & \quad \theta \in [0, \theta_{\text{ROM}}], \label{eq:prob_rom_constraint}
\end{align}
\end{subequations}
where $\left[ t_k, t_{k+1}\right]$ represents the optimization horizon, $\theta_P$ is the angular position measured from the patient, and $\theta$ and $\dot{\theta}$ are the angular position and angular velocity of the avatar's end-effector, respectively. Furthermore, $\dot{\theta}_H$ is the angular velocity reference obtained from the \emph{Hogan model}. We also note that $\theta_H$ is not explicitly used in this formulation but is implicitly involved in the computation of $\dot{\theta}_H$.

The constraint in Eq.~\eqref{eq:prob_rom_constraint} ensures that the avatar's motion respects the assigned range of motion. Additionally, the formulation of the cost function in Eq.~\eqref{eq:prob_cost} allows the autonomous avatar to minimize the control effort, which is weighted by the tunable positive parameter $\eta$. Moreover, the avatar's behavior can exhibit a spectrum ranging between closely following the patient's execution and strictly adhering to the reference motion produced by the \emph{Hogan model}, determined by the weights $\alpha_p\in \left[0,1\right]$ and $\alpha_s\in \left[0,1\right]$, respectively.

In particular, if $\alpha_p \simeq 0$, the avatar acts as a ``blind'' leader, disregarding the patient's movements and focusing solely on minimizing the velocity error with respect to the reference signal. Conversely, if $\alpha_p \simeq 1$, the avatar behaves as a perfect follower, precisely matching its position to that of the patient. For any intermediate value of $\alpha_p$, the avatar's motion will be a balance between tracking the human motion and following the reference velocity of the \emph{Hogan model}, allowing for the implementation of a wide spectrum of behaviors ranging from leader to follower, and vice versa.
Moreover, the condition $\alpha_p + \alpha_s = 1$ must be met to obtain a meaningful avatar behavior and keep it in the range between the patient execution and the reference \emph{Hogan model} behavior.

Appropriate tuning of these parameters is crucial for a successful rehabilitation strategy. Throughout the therapy, $\alpha_p$ and $\alpha_s$ need to be adjusted to adapt the avatar's motion to the patient's progress (or regression).

\subsection{Interpretable adaptation via ability index}
To evaluate the patient's performance, we first introduce an index of smoothness $J$ as a measure of the smoothness of the patient's motion. Formally, it is defined as in \cite{Hogan2009}:
\begin{equation}
    J = \dfrac{T^3}{\left( \max \dot{\theta} \right)^2} \int_{0}^{T} \dddot{\theta}^2 \; dt,
\label{eq:smoothness}
\end{equation}
where $T$ is the duration of the movement. The index $J$ quantifies the shape of the movement, independent of its duration and amplitude. Using this smoothness index, we can define another useful metric to quantify the patient's progress: the \emph{ability index} $I_A$. This index is defined as:
\begin{equation}
    I_A(n) = \dfrac{J_H(n)}{J_P(n)} \qquad \in [0,1],
\label{eq:ability_index}
\end{equation}
where $J_P$ is the smoothness index of the patient in session $n$, calculated using (\ref{eq:smoothness}), and $J_H$ is the ideal smoothness index provided by the \emph{Hogan model} for session $n$. Assuming that $J_H$ represents the optimal smoothness for a given movement trajectory, $I_A = 1$ indicates a perfect match where the patient's smoothness equals the reference motion's smoothness ($J_P = J_H$). Otherwise, as the patient's trajectory deviates further from the reference, the index tends towards zero ($I_A \rightarrow 0$).

Considering iteration $n$ during a session, the parameters $\alpha_p$ and $\alpha_s$ are then defined based on the change in the index $I_A$ between two consecutive sessions. Specifically:
\begin{equation}
\begin{cases}
    \alpha_p\left(n\right) = \alpha_p\left(n-1\right) + [I_A\left(n-1\right) - I_A\left(n-2\right)] \\
    \alpha_s\left(n\right) = 1 - \alpha_p\left(n\right)
\end{cases}
\label{eq:adaptive_alpha_p}
\end{equation}
Here, the parameter $\alpha_p$ for session $n$ is increased (or decreased) by an amount proportional to the change in the ability index $I_A$ from the previous session. The parameter $\alpha_s$ is then set as the complement of $\alpha_p$.

The adaptation mechanism in Eq. (\ref{eq:adaptive_alpha_p}) has a direct clinical interpretation. The parameter $\alpha_s$ represents the ``therapeutic challenge level'': higher values indicate the system is pushing the patient toward healthier movement patterns, while lower values indicate accommodation of current limitations. This provides clinicians with an intuitive metric. When $\alpha_s < 0.3$, the patient may need additional intervention; when $\alpha_s > 0.7$, the patient is approaching healthy movement patterns.These thresholds were derived from preliminary observations and require validation in clinical trials. The rate of change $\Delta I_A$ indicates rehabilitation velocity, with positive values showing improvement and negative values signaling potential fatigue or regression requiring clinical attention.

\section{Virtual Reality Implementation}
\label{sec:vr_software_design}

\begin{figure}[b!]
    \centering
    \includegraphics[width=0.45\textwidth]{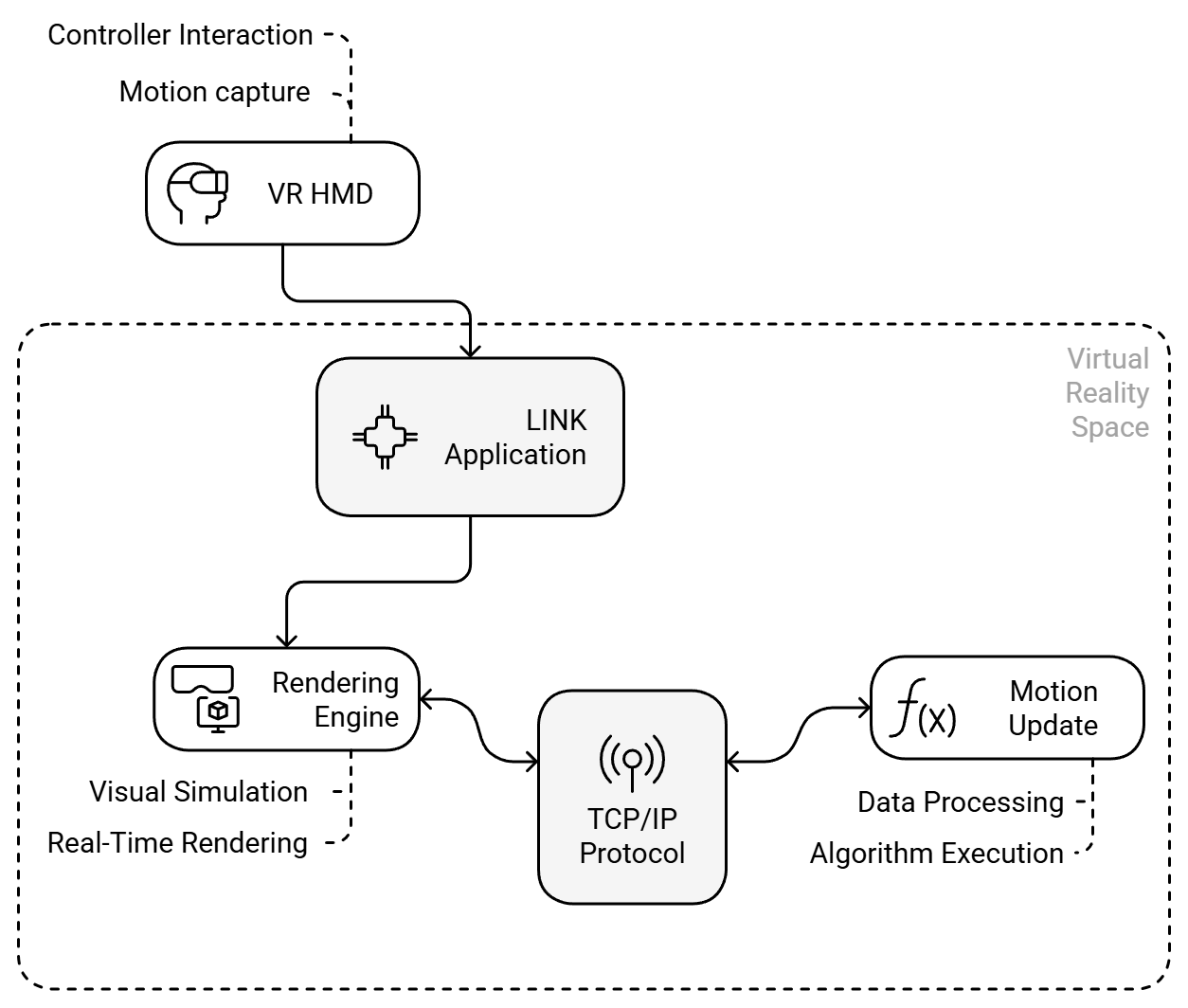}
    \caption{\textbf{System Architecture.} Conceptual diagram of the system architecture. Hand movements are tracked using Meta Quest controllers and transmitted via Quest Link to the rendering engine (Unreal Engine) hosted on a dedicated computer. A TCP/IP communication protocol allows data exchange between the Unreal Engine \emph{front-end} and a MATLAB \emph{back-end} responsible for avatar motion computation. This setup facilitates flexible prototyping of control algorithms}
    \label{fig:SystemArchitecture}
\end{figure}

The Virtual Reality (VR) system architecture presented in this work was implemented using Unreal Engine (version 5.3.2), a leading cross-platform game engine widely adopted in the industry.

Specialized devices like Head-Mounted Displays allow for effective tracking of the patient's movements, making them active participants in the virtual world as they perform various rehabilitation exercises. Key features of this VR implementation include a \textit{responsive} and \textit{immersive} environment. Responsiveness means the environment reacts in real-time as the user explores or interacts with virtual objects. Immersion aims to make the user feel present in the non-physical world, achieved through elements like synchronized audio and visuals. 

\begin{figure}[tb!]
	\centering
	\includegraphics[width=0.45\textwidth]{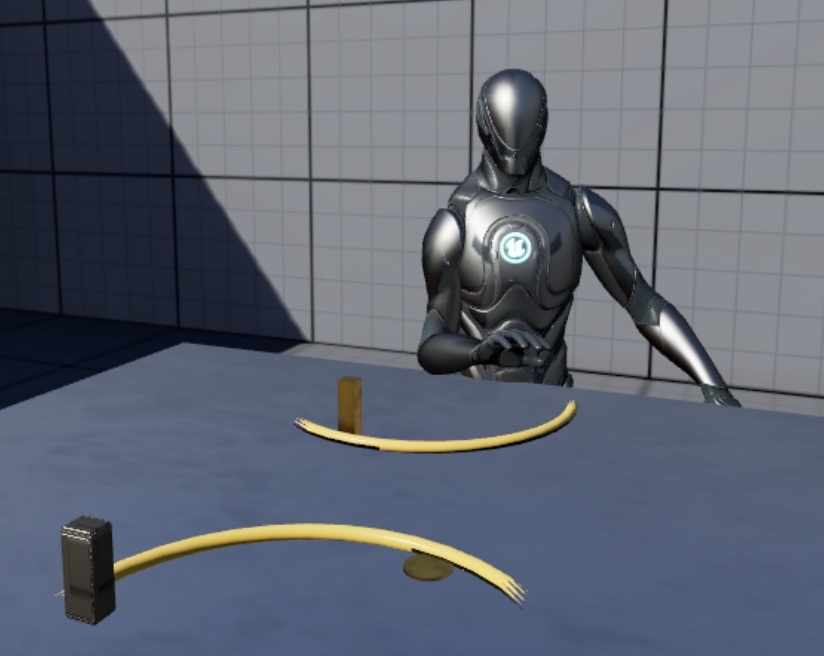}
	\caption{\textbf{User interface in virtual reality.} In the virtual environment, the user is prompted to grab the object on the right, follow the yellow trace, and reach the target location. Simultaneously, an autonomous avatar powered by our optimal control strategy on the other side of the table performs the same exercise, adapting its behavior (according to the methodology presented in Section \ref{sec:architecture_avatar}) to guide human subject physiotherapy session via visual feedback.}
	\label{fig:vr_interface}
\end{figure}

Our system architecture, schematically depicted in Fig. (\ref{fig:SystemArchitecture}), utilizes the TCP/IP protocol for seamless data exchange between the Unreal Engine \emph{front-end} and the MATLAB \emph{back-end}. This communication channel updates the avatar's motion based on the computation of the cognitive process made by the Control Architecture detailed in Section~\ref{sec:architecture_avatar}. This architectural choice offers flexibility for future Control Architecture development with minimal impact on platform deployment. Moreover, TCP/IP facilitates offloading data storage and algorithms to a dedicated machine, potentially enhancing security and reducing the computational load on the patient's system. 
Within the immersive VR environment depicted in Fig. (\ref{fig:vr_interface}), we utilize measurements from the Meta Quest controllers to track the user's end-effector positions. These measurements are transmitted to the \emph{back-end} computation module as string messages containing: (i) 3D position data and (ii) a timestamp.

\begin{figure*}[t!]
	\centering
	\includegraphics[width=0.9\textwidth]{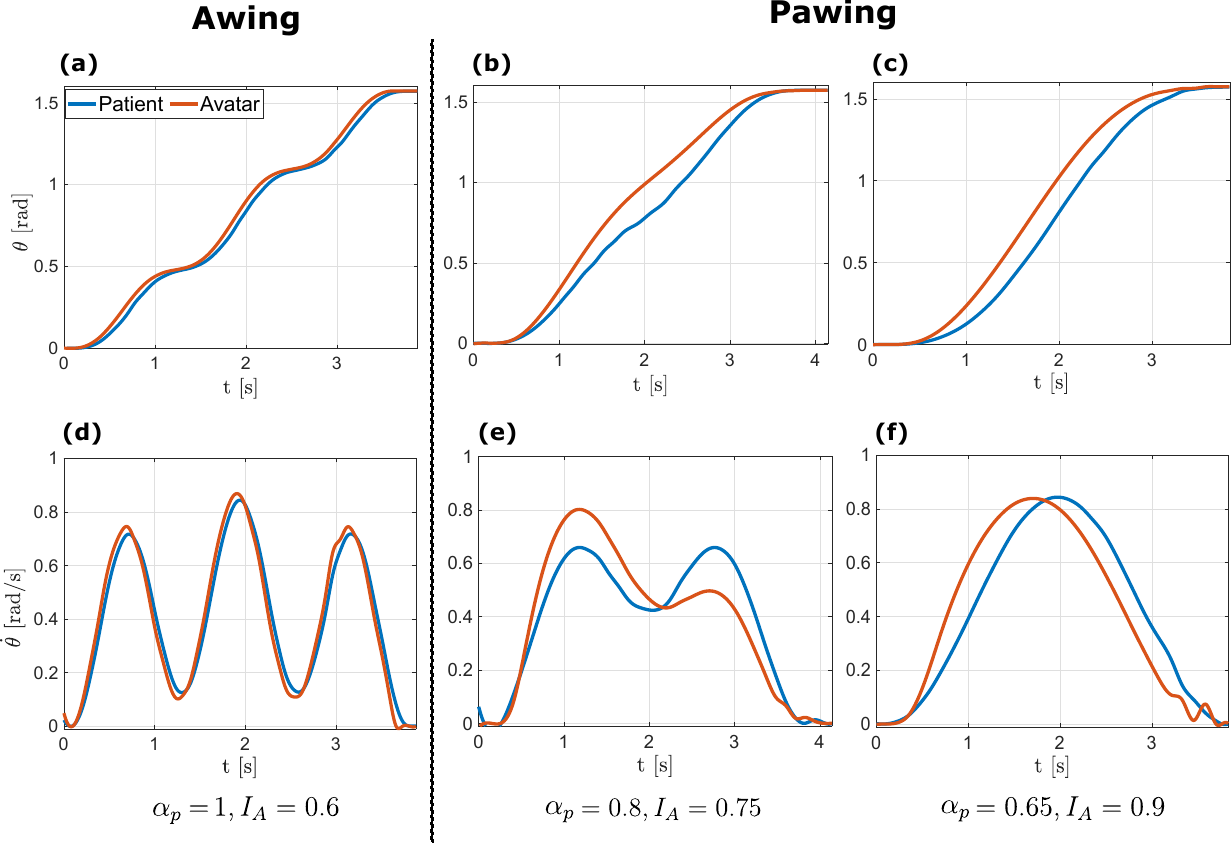}
	\caption{\textbf{Mathematical simulation of the rehabilitation process assisted by the avatar.} Mathematical simulation of the rehabilitation process showing the patient (in blue) assisted by the avatar (in red) during the phases of Awing and Pawing. \textbf{(a)-(d)} Position and velocity time series of the patient (in red) and the avatar (in blue) are reported for an Awing trial having the maximum angle set at $90$ degrees. The patient was simulated to have two hesitations during the task ($m=3$ in Eq.~\eqref{eq:patient_simulation}). \textbf{(b)-(e)} Position and velocity time series are reported for a Pawing trial that goes from $0$ to $90$ degree. The patient was simulated to have only one hesitation ($m=2$ in Eq.~\eqref{eq:patient_simulation}). \textbf{(c)-(f)} Position and velocity time series are reported for a Pawing trial simulating the patient with no hesitation ($m=1$ in Eq.~\eqref{eq:patient_simulation}).
	Furthermore, for each trial the weight of the avatar's control law $\alpha_p$ and the index of ability $I_A$ are reported showing that while one decreases the other one increases simulating the progresses of the patient.}
	\label{fig:indago_sim}
\end{figure*}

To evaluate the suitability of TCP/IP for this application, we first estimate potential error by assuming direct access to end-effector data within Unreal Engine (i.e., neglecting controller-to-render latency). We further assume a maximum patient movement speed of $1\,\text{m/s}$. Under these conditions, we analyze the impact of communication delay on the avatar's responsiveness. For instance, at a speed of $1\,\text{m/s}$, a $20\,\text{ms}$ communication delay would introduce a $20\,\text{mm}$ positional error relative to the ideal autonomous avatar movement. In our specific implementation, running the complete \emph{system architecture} on a Local Area Network (LAN), we measured an average round-trip communication delay of approximately $2\,\text{ms}$. This translates to a potential positional error of only $4\,\text{mm}$, which is considered acceptable for our application.

\section{Experimental design and validation}
\label{sec:validation}
The ability of the proposed avatar to adapt its behavior in different interaction scenarios was assessed experimentally. Specifically, the avatar is required to either lead or follow the human partner by adjusting its control parameters according to the partner’s degree of motor impairment. 
Initially, the control strategies were validated through mathematically simulations of patient motion profiles representing different levels of stroke severity. Furthermore, preliminary experimental trials with healthy participants were carried out to evaluate the feasibility and usability of the system under real and plausible motion conditions.

\subsection{Synthetic data and numerical validation}
As first validation step, a series of synthetic motion profiles were built to mimic a typical point-to-point motion, as in \cite{Hogan2009}. With this aim, three terms were superposed: \textit{i)} a linear ramp, \textit{ii)} a rescaled sinusoid with a period equal to the movement duration, and \textit{iii)} a rescaled sinusoid with a period equal to an integer $m$ divisor of the movement duration. 
Formally, the resulting function was:

\begin{equation}
\label{eq:patient_simulation}
\begin{split}
    \theta_P(t) = A \Bigg( & \dfrac{t}{T} - \sin \left(\dfrac{2 \pi t}{T}\right) \dfrac{1-b}{2 \pi} \\
    & - \sin \left( \dfrac{2m \pi t}{T} \right) \dfrac{b}{2m \pi} \Bigg),
\end{split}
\end{equation}

\noindent where $A$ is the movement amplitude, $T$ is the movement duration, and $b$ is a constant defining the deviation from a smooth cycloidal function. The constant $m$ defines the shape of the movement. Specifically, with $m=1$ the movement has a single-peak velocity profile, otherwise with $m>1$ the movement results in a multi-peaked velocity profile with $m$ peaks. Conceptually, the presence of these peaks in the velocity aims at simulating a series of hesitations in the movement typically exhibited by individuals with impairments.

To test our platform, we simulate the motion of a patient under therapy using function \eqref{eq:patient_simulation}.
Specifically, in the Awing phase, the function (\ref{eq:patient_simulation}) was parameterized with $m=3$ and $b=0.8$ in order to simulate two patient's hesitations (trajectory with three peaks). 
The avatar's parameters in (\ref{eq:inner_dynamics}) were set as $K=0.64, \, B=0.4, \, I=0.014$ as done in \cite{Hogan1984}, while the controller was tuned to be a pure follower ($\alpha_p = 1, \, \alpha_s = 0$). In this phase, as explained in Section \ref{sec:rehabilitation_platform}, the avatar does not influence the patient's movement but only follows it in order to evaluate their index of smoothness $J$ and compare it with the index of smoothness $J_H$ computed by the \emph{Hogan model} along the same trajectory. Quantitatively, with this setting, we have $J_P = 11$ and $J_H = 6.58$ resulting in $I_A = 0.6$. In the next phase of Pawing the progresses of the patient were simulated in two exercises. In the first exercise the patient was simulated to be slightly improved setting $m=2$ and $b=0.5$ (one hesitation, trajectory with two peaks). According to (\ref{eq:adaptive_alpha_p}) the avatar was set as $\alpha_p = 0.8$ and $\alpha_s = 0.2$ showing a behavior that is a trade-off between the reference coming from the \emph{Hogan model} and the patient's motion. Quantitatively, we have $J_P = 8.7$ and so $I_A = 0.75$. Once again, according to (\ref{eq:adaptive_alpha_p}) in the next exercise $\alpha_p = 0.65$ and $\alpha_s = 0.35$, while the patient was simulated with $m=1$ and $b=0.1$ (no hesitation, one single peak). In this last exercise, $J_P = 7.35$ and $I_A=0.9$ simulate that the patient has reached a healthy profile. The result of such procedure is depicted in Fig. (\ref{fig:indago_sim}).

\subsection{Preliminary experimental trials}
\label{sec:experimental_trial}
All participants provided written informed consent in accordance with the Declaration of Helsinki prior to participation, after being fully informed about the study objectives, procedures, and their right to withdraw at any time. Given the non-invasive nature of the study involving healthy volunteers performing standard motor tasks, the research was conducted following the Declaration of Helsinki ethical principles. All data were collected and analyzed anonymously, with no personally identifiable information retained.

To evaluate the efficacy of the control strategies, we asked five healthy participants to perform the rehabilitation trials under three experimental conditions using the VR platform of Sec.\ref{sec:vr_software_design}. In the first condition (Solo), the participants were instructed to move their arm along the prescribed trajectory without the presence of the avatar.
In the second condition (Pawing), the participants were asked to follow the avatar, which was set in Pawing mode with a fixed parameter value of $\alpha_s = 0.8$ throughout the trials. Finally, in the last condition (adaptive-Pawing), the participants were instructed to follow an adaptive avatar. In this case, the avatar was initially set with $\alpha_s = 0.5$, which was subsequently and automatically adjusted across the trials according to the ability index $I_A$ as defined in Eq. (\ref{eq:adaptive_alpha_p}). Each participant completed a total of eight repetitions for each condition and the acquired signals were filtered by means of a zero-phase low pass Butterworth filter with cutoff to $2$Hz to rule out the effects of measurement uncertainties and delays in the ability index computation.

\begin{figure}[t!]
	\centering
    \begin{tikzpicture}[remember picture]
        \node (plot) at (0,0) {
            \includegraphics[width=0.445\textwidth]{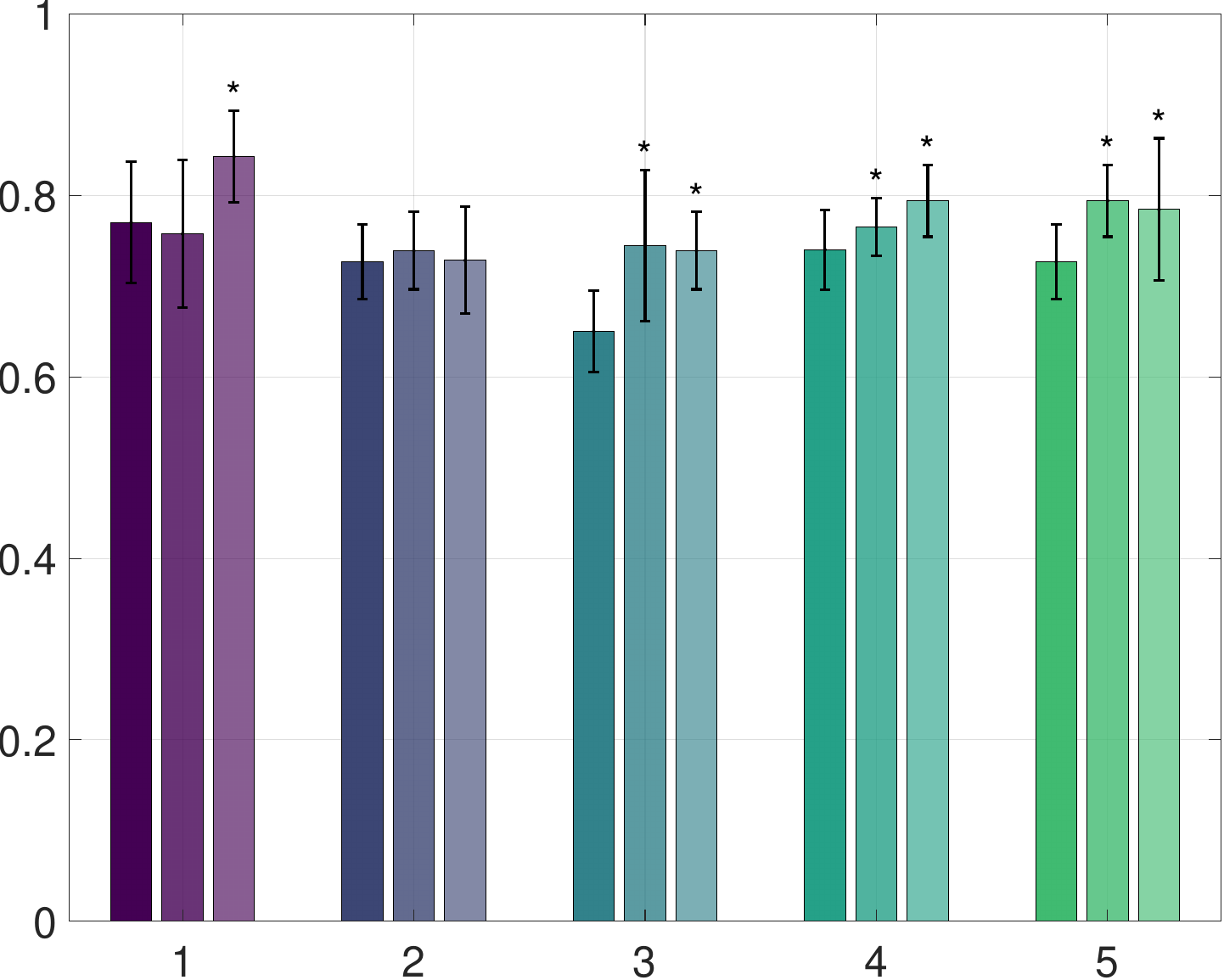}
        };
        \node[below] at (plot.south) [yshift=0.5em] {
            Subject ID
        };
        \node[] at (plot.west) [xshift=-0.5em, yshift=0.5em] {
            $I_A$
        };
    \end{tikzpicture}

    \caption{\textbf{Ability index of the rehabilitation process with and without assistance of the avatar per subject.} This histogram illustrates the ability index (calculated as in Eq. \eqref{eq:ability_index}) for the last six experimental trials to focus on steady state performance for each of the five participants, with error bars representing the standard deviation. This specific trial range highlights the steady-state performance. For each participant, the first bar represents the Solo condition, the second bar depicts the Pawing condition with  $\alpha_s = 0.8$ and the third bar refers to the adaptive-Pawing condition, where the artificial physiotherapist guided the exercise using adaptive parameters ($\alpha_p$ and $\alpha_s$) computed according to Eq. \eqref{eq:adaptive_alpha_p}. Asterisks on top of the Pawing and adaptive-Pawing indicates statistical significance ($p < 0.1$), computed via a paired t-test against Solo condition.}
    \label{fig:mean_ability_index}
\end{figure}

\begin{figure*}[t!] 
    \centering

    \begin{subfigure}[b]{0.3\textwidth}
        \centering
        \begin{tikzpicture}[remember picture]
            \node (plot) at (0,0) {
                \includegraphics[width=\textwidth]{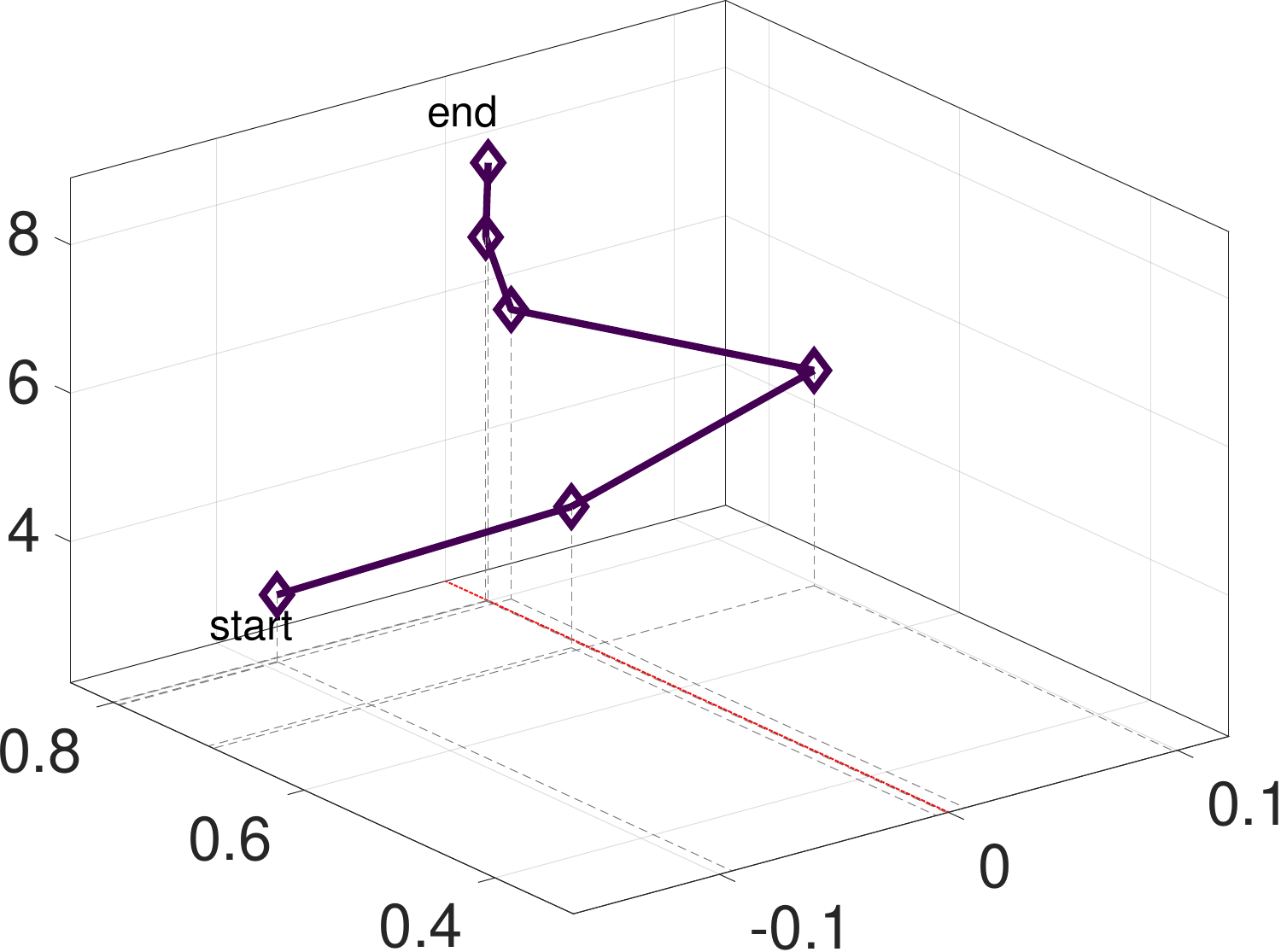}
            };
            \node[below] at (plot.south) [xshift=4.5em, yshift=0.8em] {
                $\Delta I_A$
            };
            \node[below] at (plot.south) [xshift=-4.5em, yshift=0.8em] {
                $\alpha_p$
            };
            \node[left] at (plot.west) [xshift=0.4em] {
                $n$
            };
        \end{tikzpicture}
        \caption{}
        \label{fig:ADAPT_1}
    \end{subfigure}
    \hspace{-12em}
    \hfill
    \begin{subfigure}[b]{0.3\textwidth}
        \centering
        \begin{tikzpicture}[remember picture]
            \node (plot) at (0,0) {
                \includegraphics[width=\textwidth]{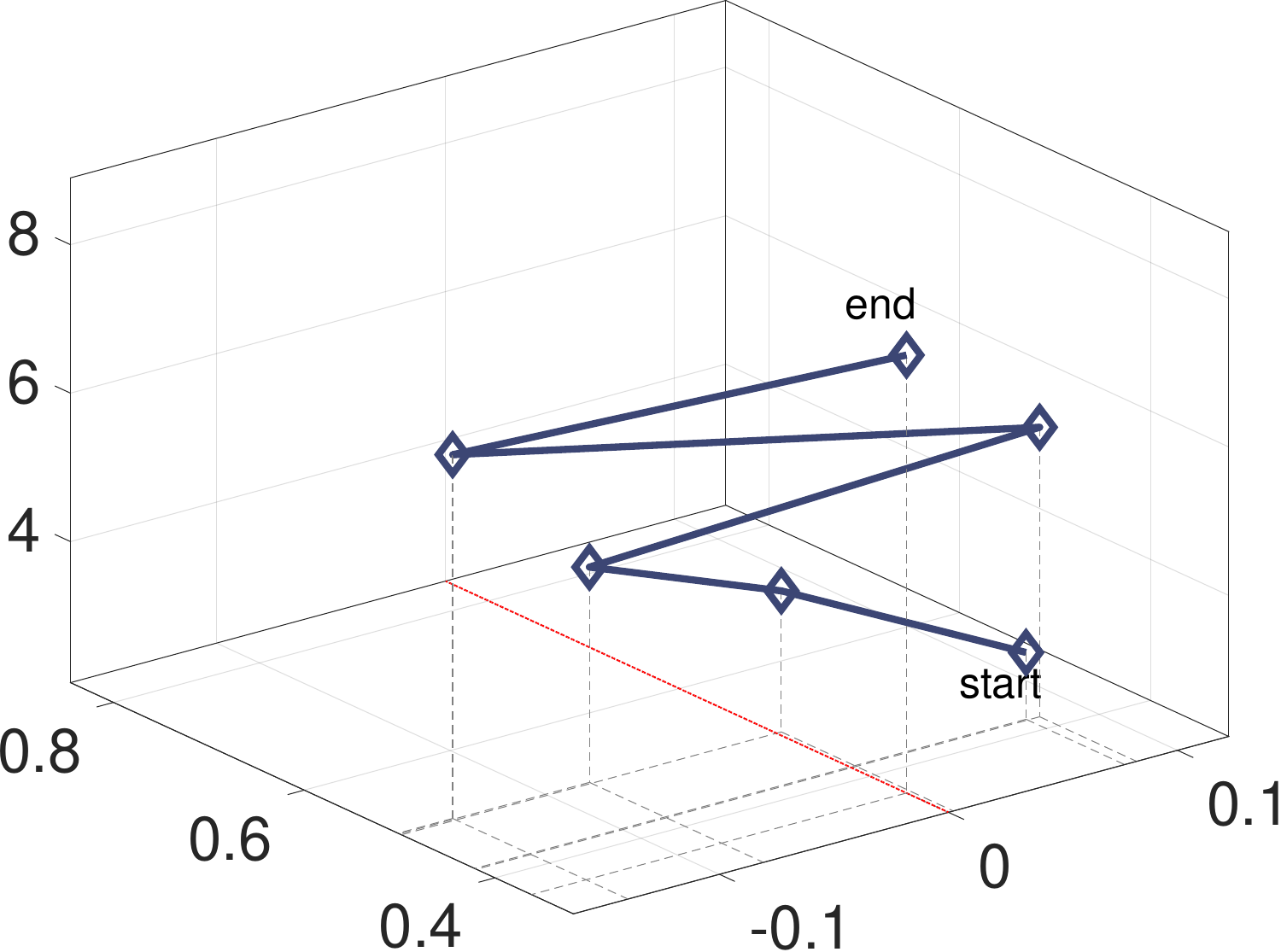}
            };
            \node[below] at (plot.south) [xshift=4.5em, yshift=0.8em] {
                $\Delta I_A$
            };
            \node[below] at (plot.south) [xshift=-4.5em, yshift=0.8em] {
                $\alpha_p$
            };
            \node[left] at (plot.west) [xshift=0.4em] {
                $n$
            };
        \end{tikzpicture}
        \caption{}
        \label{fig:ADAPT_2}
    \end{subfigure}
    \hspace{-12em}
    \hfill
    \begin{subfigure}[b]{0.3\textwidth}
        \centering
        \begin{tikzpicture}[remember picture]
            \node (plot) at (0,0) {
                \includegraphics[width=\textwidth]{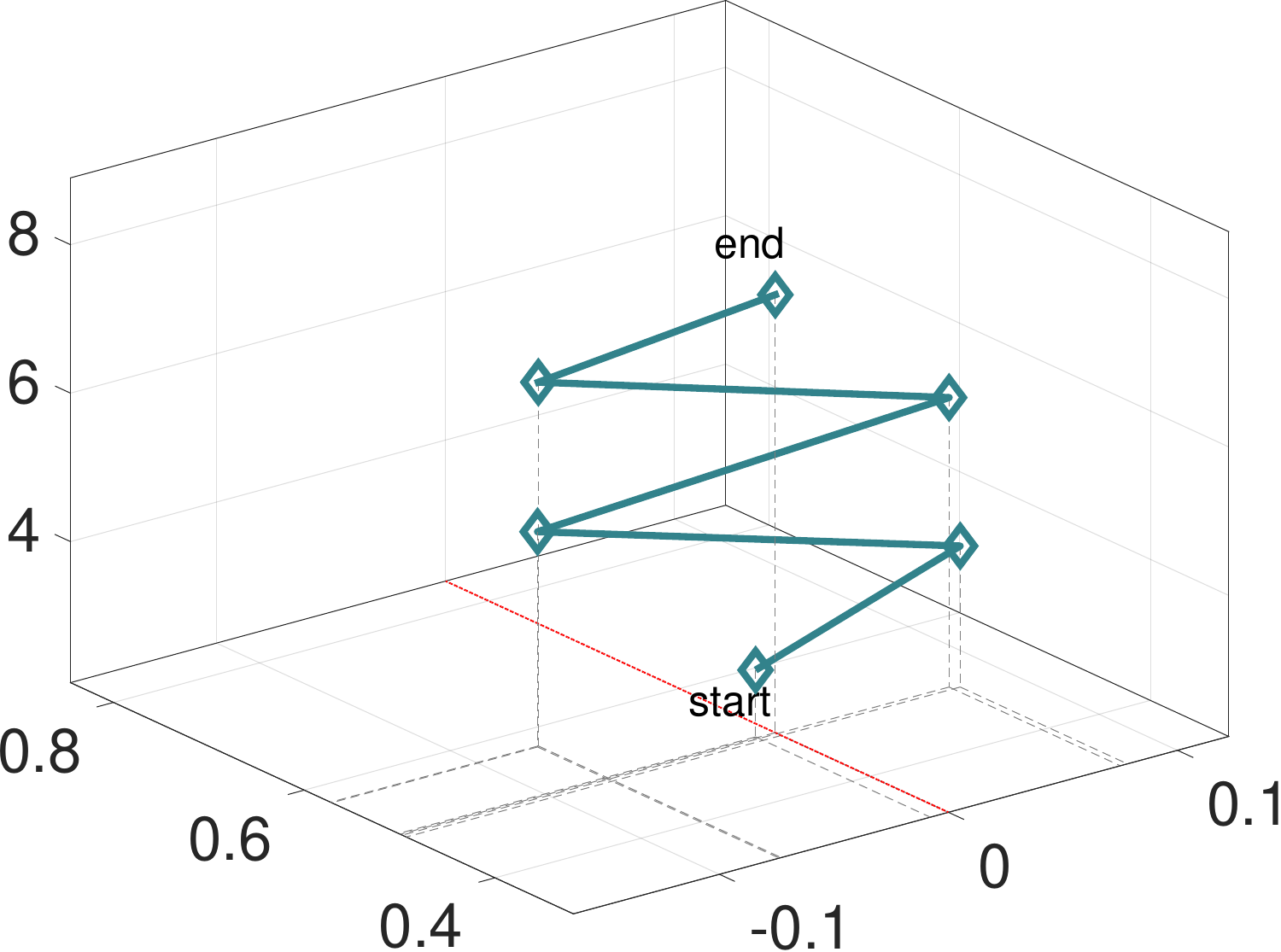}
            };
            \node[below] at (plot.south) [xshift=4.5em, yshift=0.8em] {
                $\Delta I_A$
            };
            \node[below] at (plot.south) [xshift=-4.5em, yshift=0.8em] {
                $\alpha_p$
            };
            \node[left] at (plot.west) [xshift=0.4em] {
                $n$
            };
        \end{tikzpicture}
        \caption{}
        \label{fig:ADAPT_3}
    \end{subfigure}
    \hspace{1em}

    \vfill
    
    \begin{subfigure}[b]{0.3\textwidth}
        \centering
        \hspace{-2.3em}
        \begin{tikzpicture}[remember picture]
            \node (plot) at (0,0) {
                \includegraphics[width=\textwidth]{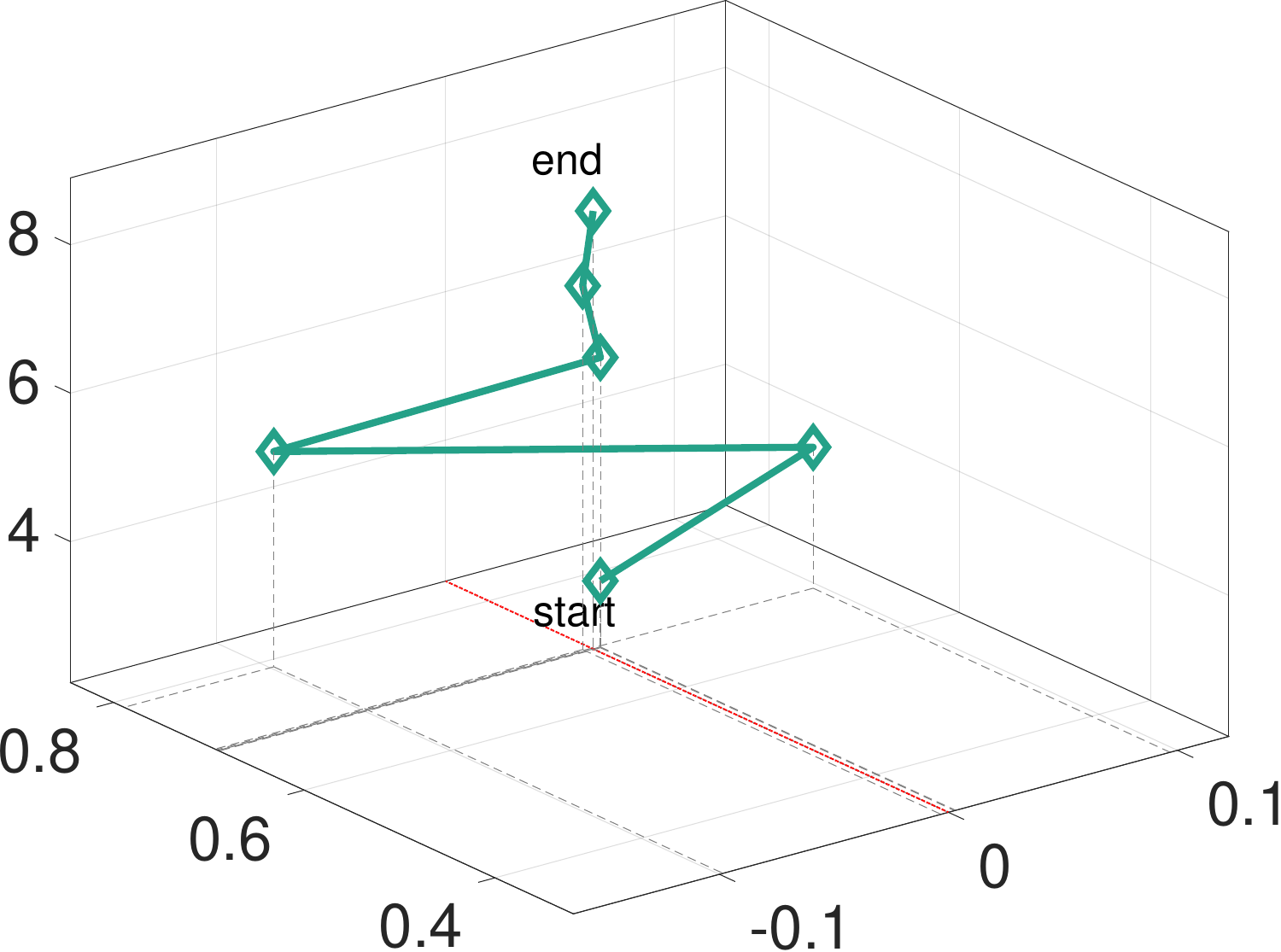}
            };
            \node[below] at (plot.south) [xshift=4.5em, yshift=0.8em] {
                $\Delta I_A$
            };
            \node[below] at (plot.south) [xshift=-4.5em, yshift=0.8em] {
                $\alpha_p$
            };
            \node[left] at (plot.west) [xshift=0.4em] {
                $n$
            };
        \end{tikzpicture}
        \caption{}
        \label{fig:ADAPT_4}
    \end{subfigure}
    \begin{subfigure}[b]{0.3\textwidth}
        \centering
        \begin{tikzpicture}[remember picture]
            \node (plot) at (0,0) {
                \includegraphics[width=\textwidth]{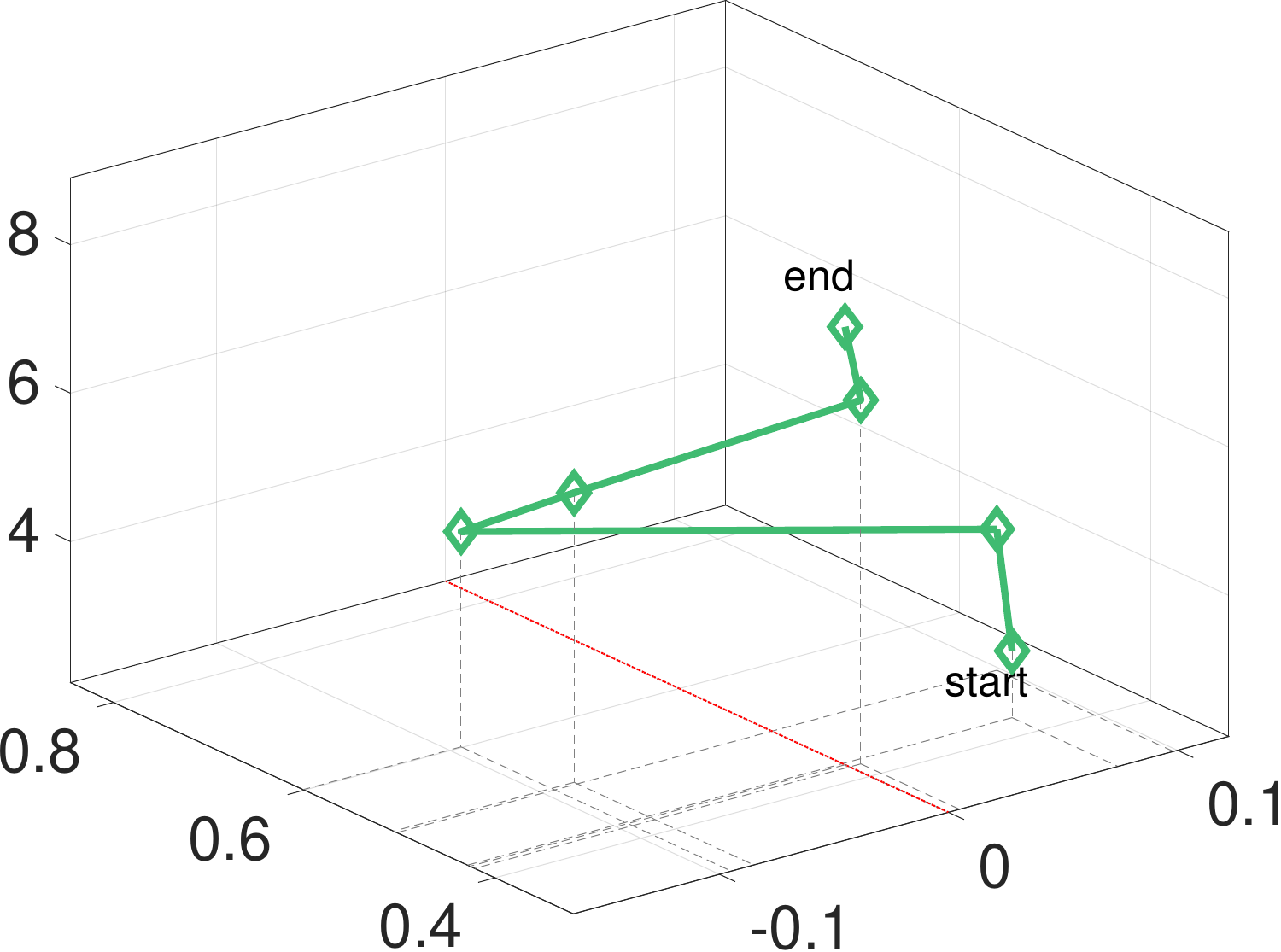}
            };
            \node[below] at (plot.south) [xshift=4.5em, yshift=0.8em] {
                $\Delta I_A$
            };
            \node[below] at (plot.south) [xshift=-4.5em, yshift=0.8em] {
                $\alpha_p$
            };
            \node[left] at (plot.west) [xshift=0.4em] {
                $n$
            };
        \end{tikzpicture}
        \caption{}
        \label{fig:ADAPT_5}
    \end{subfigure}
    \hfill
    \caption{\textbf{Individual subject charts.}  Each subplot (a-e) represents a different subject, showing their progression in a 3D space defined by avatar adaptation rate ($\alpha_p(n)$), individual ability index variation in two consecutive trials ($\Delta I_A = I_A(n) - I_A(n-1)$), and trial number ($n$). The 'start' point marks the beginning of platform adaptation at $n = 3$, while the 'end' point marks the conclusion of the session for each subject. The red dashed line indicates the plane of no variation in the subject ability index between two consecutive trials ($\Delta I_A = 0$). Subjects $1, 4$ and $5$ (panels (a), (d) and (e), respectively) converge towards the $\Delta I_A=0$ plane indicating performance stabilization and successful motor learning. Subjects 2 and 3 (panels (b), (c)) exhibit sustained oscillations, suggesting inconsistent skill retention and need for clinical intervention adjustment.}
    \label{fig:parameter_variation}
\end{figure*}
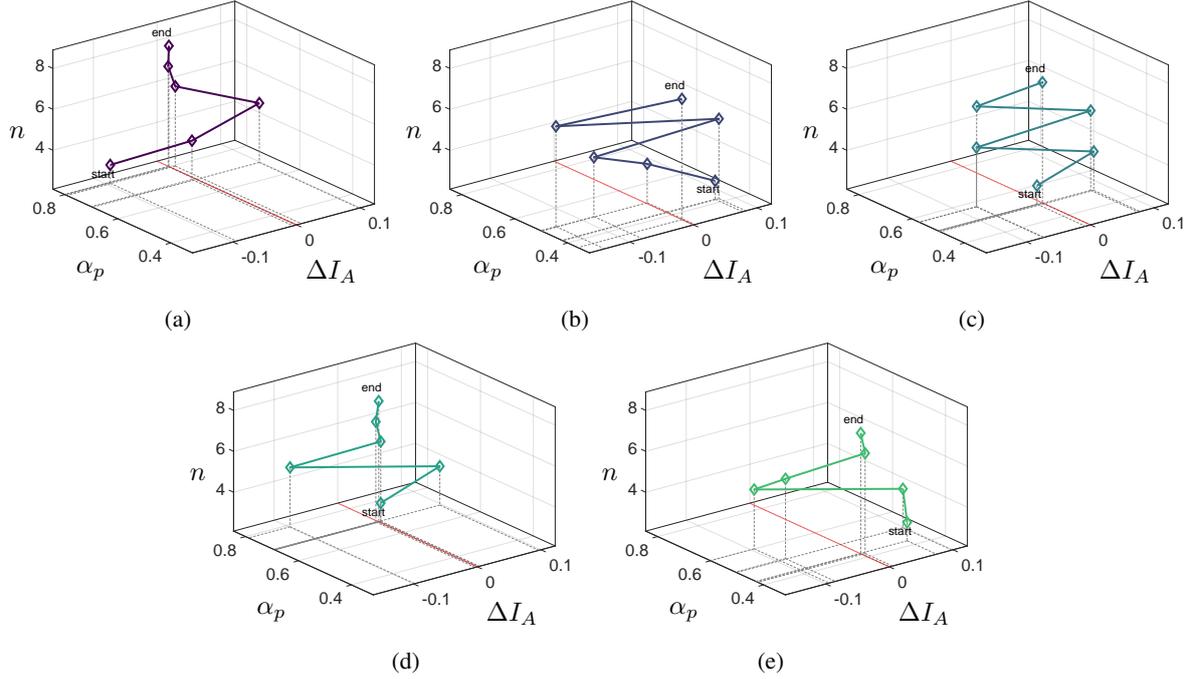

Fig. (\ref{fig:mean_ability_index}) reports mean and 1-sigma reproducibility of the ability index across the last six trials for all participants in the Solo, Pawing and adaptive-Pawing conditions to capture steady state performance. The results indicate significant inter-subject non-reproducibility  in performance. Notably, participant $3$ exhibited a marked increase in the ability index under the Pawing and adaptive-Pawing conditions respect to the Solo condition, suggesting a positive effect of the avatar guidance.
For all the participants, the average performance in the Pawing and adaptive-Pawing condition was never lower than in the Solo condition. Furthermore, compatible results were observed between Pawing and adaptive-Pawing condition. This outcome is consistent with the fact that all participants were healthy subjects.

Individual adaptation trajectories for each subject in the adaptive-Pawing condition are shown in Fig.~\ref{fig:parameter_variation}. The three-dimensional space, defined by trial number ($n$), platform adaptation rate ($\alpha_p$) and change in ability index ($\Delta I_A = I_A(n) - I_A(n-1)$), reveals subject-specific learning patterns with direct clinical significance.  Subjects 1, 4, and 5 (Fig.~\ref{fig:parameter_variation}(a), (d), (e)) demonstrate successful adaptation as their trajectories converge toward the $\Delta I_A = 0$ plane as trials progress, indicating performance stabilization at consistent ability levels ($I_A(n) \approx I_A(n-1)$). This convergence reflects successful motor learning and skill retention in response to the avatar's adaptive challenge.
In contrast, subjects 2 and 3 (Fig.~\ref{fig:parameter_variation}(b), (c)) exhibit sustained fluctuations in $\Delta I_A$, frequently crossing the $\Delta I_A = 0$ plane. Clinically, this oscillatory pattern may indicate inconsistent skill retention or suboptimal challenge level, signaling need for intervention adjustment. These individual trajectories provide objective, trial-by-trial metrics for monitoring progress, diagnosing stagnation, and personalizing therapeutic strategies beyond subjective patient reports.

\section{Conclusions}
\label{sec:conclusions}
This paper presented a control-based rehabilitation platform designed to support post-stroke survivors during the long-term outpatient phase. The system enables patients to perform structured motor exercises at home, assisted by an adaptive autonomous virtual avatar in a virtual reality environment, while clinicians can monitor progress remotely. At the core of the platform lies an optimal control architecture that drives the avatar’s motion in coordination with the patient’s and a set of tools for quantitative evaluation of subject performance. The control input is generated by minimizing a cost functional that balances two objectives: synchronizing with the patient's current trajectory and guiding them toward an ideal reference motion derived from the Hogan minimum-jerk model. Crucially, the control framework’s primary strength is its inherent adaptability: it doesn't rigidly enforce a fixed target but embeds a parameter adaptation mechanism that updates the weights of the cost function across sessions, based on a smoothness-based ability index \( I_A \). This mechanism is precisely what allows the platform to adapt the level of assistance and guidance to the patient's current motor capabilities and rehabilitation progress.

Experimental and simulation results demonstrate the avatar’s ability to co-adapt with the user and gradually reinforce smoother, more controlled movements over time. The reference trajectory not only serves as a benchmark for healthy motion but also enables a quantitative assessment of the patient’s motor quality throughout the rehabilitation process.
This patient-centric approach ensures that the rehabilitation challenge remains appropriate while powering the our adaptive optimal control.
While the framework exhibits promising behavior, important limitations must be acknowledged. Only a one-degree-of-freedom reaching task has been validated, and validation was conducted exclusively with healthy participants. Clinical validation with stroke patients remains essential to assess efficacy and safety.

Future work will focus on extending the proposed approach to multi-joint and full-arm models, performing rigorous analysis of convergence and boundedness under time-varying cost functions, and validating the platform in clinical settings with stroke patients. 

\section*{Acknowledgements}
The authors thank Dr. Giovanni D'Addio from the Istituti Clinici Scientifici Maugeri IRCCS, Telese Terme, Italy, for his insightful contributions to formulating the initial methodology for this work.

\end{document}